\documentclass[pre,twocolumn,epsfig,eqsecnum,superscriptaddress]{revtex4} 
\usepackage{graphicx,subfigure} 
\usepackage{dcolumn} 
\usepackage{graphicx} 
\usepackage{bm} 
\usepackage{amssymb} 

\begin{document} 
\title{Computer simulations on the sympatric speciation modes for the 
Midas cichlid species complex} 
\author{K. Luz-Burgoa} 
\author{S. Moss de Oliveira} 
\author{J. S. S\'a Martins} 
\address{Instituto de F\'{\i}sica, Universidade Federal 
Fluminense, Campus da Praia Vermelha, Boa Viagem, Niter\'oi, 24210-340, RJ, 
Brazil.} 
\email{karenluz@if.uff.br} 
 
\date{\today} 
 
\begin{abstract}
Cichlid fishes are one of the best model system for the study of 
evolution of the species. Inspired by them, in this paper 
we simulated the splitting of a single species into two separate ones via
random mutations, with both populations living together in sympatry, 
sharing the same habitat.  
We study the ecological, mating and genetic conditions needed to reproduce 
the polychromatism and polymorphism of three species of the Midas Cichlid 
species complex. 
Our results show two scenarios for the {\it A. Citrinellus} speciation 
process, one with and the other without disruptive natural selection. 
In the first scenario, the ecological and genetic conditions are sufficient
to create two new species, while in the second the mating and genetic 
conditions must be synchronized in order to control the velocity of genetic 
drift.
\end{abstract} 

\maketitle

\section{Introduction}

The evolution of a single population into two or more species without
prevention of gene flow through geographic segregation is known as
sympatric speciation \cite{mayr,bush,futuyma}. Cichlid fishes, {\it Amphilophus
 Zaliosus}, are one possible example of evolution by sympatric speciation 
in nature \cite{marta,seehausen}. 
The Midas cichlid species complex \cite{barlow} are distributed in the Great Lakes of 
Nicaragua as well as in several crater lakes in the area. 
Combined, they represent by far the largest biomass of any fish species in 
Nicaragua freshwaters. There are substantial morphological differences between them, for example, 
some of them have cryptic colouring, grey or brown with dark bars or spots, known as Normal Morph, or a 
conspicuous form, which lacks melanophores, resulting in brightly red, 
orange, or yellowish coloured fish, Gold Morph. 
Cichlid fishes are also characterized by a pair of jaws in the pharyngeal 
area in addition to the oral jaws, and this key innovation is presumed to be responsible 
for their great ability to colonize new habitats and to exploit successfully a large 
diversity of trophic niches.

\begin{figure}[htbp]
\begin{center}
\includegraphics[width=7.8cm]{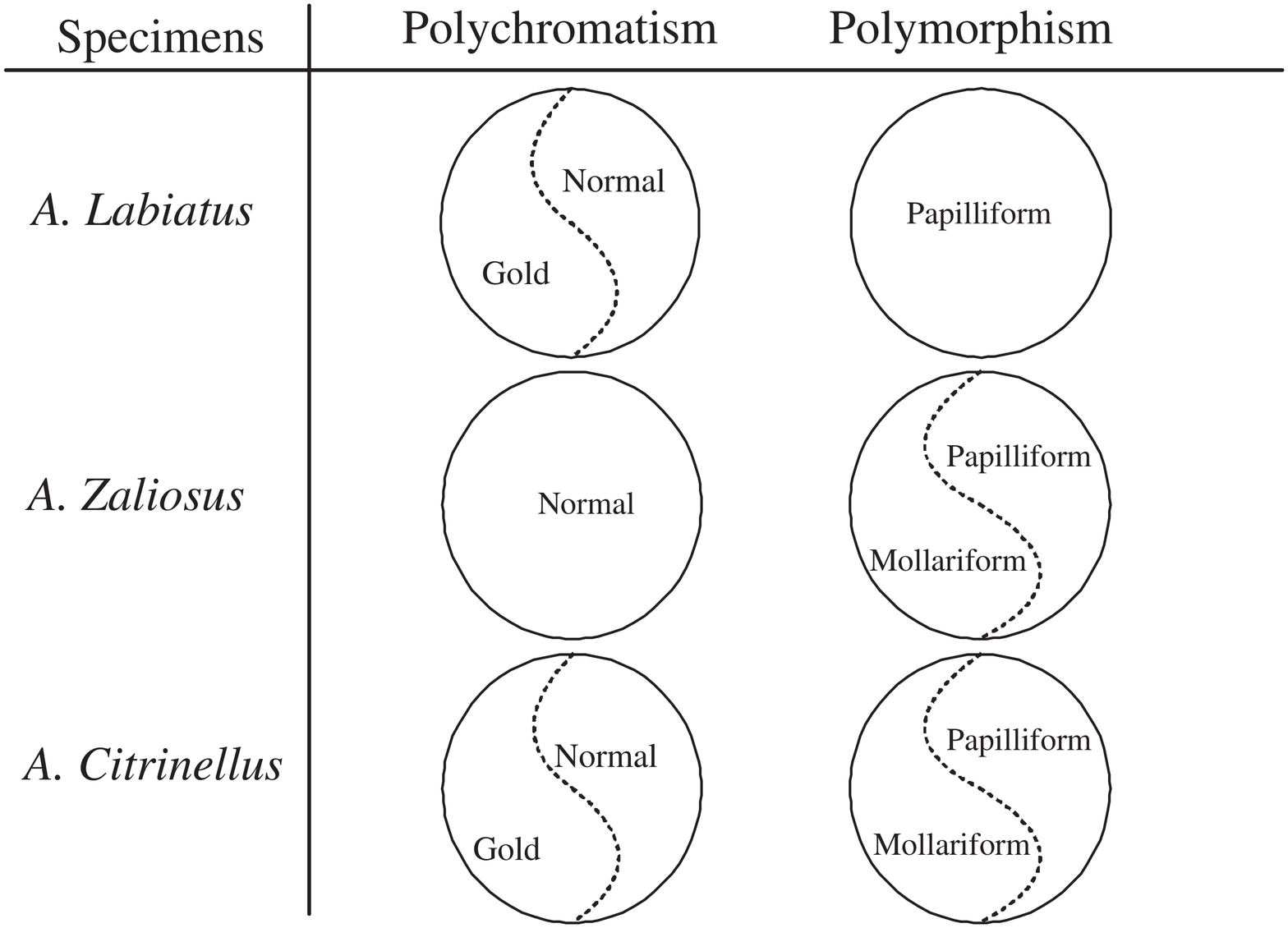}
\end{center}
\caption{\it The scheme shows the polycromatism and polymorphism 
of the three species of the Midas cichlid species complex 
 from the lakes of Nicaragua described.}
\label{fig:conjuntos}
\end{figure}

Three different species have already been recognized \cite{barlow} within the 
Midas cichlid complex: 
1.- {\it Amphilophus Labiatus}, the red devil cichlid, a fleshy-lipped species
thought to be restricted to the big lakes Lake Nicaragua and Lake Managua, 
2.- {\it A. Zaliosus}, the arrow cichlid, an elongated species that is restricted
in its distribution to one of the crater lakes, Lake Apoyo, and
3.- {\it Amphilophus citrinellus}, the Midas cichlid, 
a generalist species with very widespread distribution.
For {\it A. citrinellus}, a polychromatism has been described 
\cite{martameyer}, with normal and gold morphs. 
Strong assortative mating according to colour has been 
observed as well \cite{wilson}, both in the field and in captivity, suggesting 
that sexual selection maintains the colour polymorphism. 
Two types of pharyngeal jaws of {\it A. citrinellus} 
have been described \cite{meyer1989}, a papilliform morph with slender 
pointed teeth and a mollariform morph with thicker rounded teeth. 
These previous studies showed that there is a 
trade-off in performance: the mollariform fish are specialized 
and more efficient at eating hard diets such as snails, whereas they are less 
efficient in feeding on soft diets than the papilliform morphs and vice versa.
For {\it A. labiatus}, the same polychromatism has been described \cite{martameyer}, 
with normal and gold morphs, and only papilliform pharyngeal jaws have been documented. 
All {\it A. zaliosus} have cryptic colouring, normal morph, and are
polymorphic with papilliform and mollariform pharyngeal, Fig. \ref{fig:conjuntos}. 

According to theoretical models \cite{sara}, sympatric speciation is driven 
by disruptive frequency-dependent natural selection, caused by competition
for diverse resources \cite{kk,dd}. On the other hand, some authors have argued that 
sexual selection can also cause sympatric speciation \cite{takimoto,gavrilets}. 
Cichlids are renowned for their vast diversity of trophic morphologies and often 
extreme degree of ecological specialization \cite{meyer1989}. 
However, the sympatric occurrence of many sibling species that seem to differ 
only in colouring makes it unlikely for ecological specialization to be 
the sole mechanism of speciation in this group.
At the same time, genetic data \cite{marta} show that sympatric speciation by sexual 
selection alone is rather unlikely for the speciation case of the crater lake Apoyo.
We studied the ecological, mating and genetic conditions needed to reproduce 
the polychromatism and polymorphism of the three species of the 
Midas Cichlid species complex, Fig. \ref{fig:conjuntos}.
Our study was based on simulations of an individual-based model where 
natural selection caused by competition for diverse resources 
and sexual selection, tuned by strength parameters on two quantitative 
and independent traits, were considered. 
Our results show two scenarios for the sympatric 
speciation of {\it A. Citrinellus} species, one with and the other without disruptive 
natural selection. In the first scenario, {\it A. Zaliosus} develops jaw polymorphism 
while retaining a single colour morph, while in the second {\it A. Labiatus} develops 
polychromatism with a single jaw morph.

\section{Computational model}

The individuals' genomes are represented by three pairs of bitstrings, 
each of them consisting of a computer word of 32 bits. 
The first pair is age-structured and contains the information of when, between 1 and 32 
time steps, the individual would die 
if only genetic diseases were considered \cite{penna}. 
The second pair delineates the ability of the individual to survive 
under a competition for the available resources, representing an ecological 
trait such as the types of pharyngeal jaws in the case of cichlid fishes. 
The third pair represents a trait only related to sexual selection, a mating
trait, such as the colour of cichlid fishes. 
At the beginning of the simulation, all individuals are born with random genomes. When a female 
succeeds in staying alive until reaching a minimum reproduction age, $A$, it 
looks for a male to mate with and generates $b$ offspring every time step before dying, 
with a new choice for a mate being done at each time step.
The first pair of the offspring's genome is constructed in the following way: 
each one of the first pair of strings of the male, for instance, is broken at the 
same random position and the complementary pieces, originated from different strings, 
are joined to form two male gametes. 
One of the gametes is then randomly chosen to be passed on to the offspring. 
After that, one random bad mutation is introduced into this gamete, and the 
final result corresponds to one string of the new individual. The other string 
is constructed from the first pair of the female's strings by the same process, 
that simulates random crossover, recombination and addition of one bad mutation.

\subsection{Natural selection caused by competition}

In the present model, competition for food is related to a phenotype, $j$, 
represented by the second, non age-structured, pair of bit-strings,
which is constructed in the same way as the first pair of the individual' genome.
This phenotypic characteristic is computed by counting the number of bit 
positions where both bits are set to $1$, plus the number of dominant positions 
(chosen as $16$) with at least one of the two bits set. It will therefore 
be a number $j$ between $0$ and $32$, which we will refer to as 
the individual's phenotype. 
We call $M_j$ the mutation probability per locus of this ecological trait. 
A mutation can change the locus either from $0$ to $1$ or from $1$ to $0$. 
In order to control the population's size and introduce a competition we 
used the Verhulst factor, $V(j,t)$. We considered three intra-specific 
competitions, \cite{phase}, depending on the individual's phenotype $j$, 
each one related to a given phenotypic group: 
\begin{equation}\label{fatorall} 
V(j,t) = \left\{\begin{array}{rcl} V_1(j,t), & 0\le j<n_1;&{\rm specialist,}\\ 
V_m(j,t), & n_1\le j\le n_2;&{\rm intermediate,}\\ 
V_2(j,t), & n_2< j\le 32;&{\rm specialist.},
\end{array}\right. 
\end{equation} 
where $n_1$ and $n_2$ are two parameters of the simulation. 
At every time step, $t$, and for each individual with phenotypic 
characteristic $j$, a random real number uniformly distributed 
between $0$ and $1$ is generated; 
if this number is smaller than $V(j,t)$, the individual dies. 
For the specialist groups the competition is given by: 
\begin{equation}\label{extreme} 
V_{1(2)}(j,t) = \frac{P_{1(2)}(t)+P_m(t)}{F(j,t)}, 
\end{equation} 
where $P_{1(2)}(t)$ accounts for the population with phenotype $j<n_1$ 
($j>n_2$) at time $t$, $P_m(t)$ accounts for the population with phenotype $j\in 
[n_1,n_2]$, and $F(j,t)$ is a resource distribution.
Individuals with intermediate phenotypes ($P_m$) compete among themselves and 
also with a fraction $X$ of each specialist population.
The Verhulst factor for them is: 
\begin{equation}\label{intermediate} 
V_m(j,t) = \frac{P_m(t)+X\times\left[P_1(t)+P_2(t)\right]}{F(j,t)}, 
\end{equation} 
Eq.(\ref{extreme}) means that specialist individuals ($P_1$, $P_2$) 
compete with those belonging to the same phenotypic group 
and also with the whole intermediate population, 
but there is no competition between specialists of different groups 
because we are assuming they are specialized to some extent 
($[0,n_1)$,$(n_2,32]$) on particular resources, as is the case for 
papilliform and mollariform pharyngeal jaws in the Midas cichlid
species complex.
In equations \ref{extreme} and \ref{intermediate}, the resource 
distribution used varies according to:
\begin{eqnarray}\label{eq:gauss}
F(j,t)&=&C\times\left(1-G(j)\right), \mbox{with}\\
G(j)&=&Z\times e^{-\left(16-j\right)^2/64},\nonumber
\end{eqnarray}
where $C$ is a carrying capacity, and for all simulations it was set to $C=2\times10^5$.
The first case, $Z=0$ in Eq. (\ref{eq:gauss}), is used to simulate a scenario without 
disruptive selection on the ecological trait. For this case, all individuals have the same 
carrying capacity, Eq. (\ref{extreme}) and Eq. (\ref{intermediate}).
The second case, $Z>0$ in Eq. (\ref{eq:gauss}), is used to simulate disruptive natural 
selection with a strength $Z$. For this case, all individuals with intermediate phenotypes 
are disadvantaged, with respect to specialist individuals, according to a reversed 
gaussian, $G(k)$. 

\subsection{Sexual selection}

In the simulations, sexual selection is related to another phenotype, $k$, 
and was represented by a new pair of non age-structured bit-strings, the 
mating trait, that also obeys the general rules of crossing and recombination.
This phenotype was computed in the same way as that for the ecological trait. 
It will therefore be a number $k$ between $0$ and $32$, and we call $M_{k}$  
the mutation probability per locus of this mating trait, which can also mutate back and 
forth between $0$ and $1$.  
In order to consider assortative mating in a sympatric environment, 
we defined two phenotypic groups, one composed by the individuals that have $k\in[0,16)$ while individuals of the other have $k\in(16,32]$.
With some probability, $Y\in[0.0,1.0]$, a female with phenotype $k$ will mate with 
a male of the same phenotypic group and with probability $1.0-Y$ will mate with a 
male of the other phenotypic group, at each time step of its life. 
For instance, if a female has phenotype $k\in[0,16)$ and a random real number, 
$r$, uniformly distributed between $0$ and $1$, is tossed that is smaller 
than $Y$, it selects its mate among $N_m$ males of the same phenotypic group 
$k\in[0,16)$ by picking the one with the smallest phenotype value $k$. 
If a female has phenotype $k\in[0,16)$ and the random real number, $r$, is 
larger than $Y$, it chooses a partner from the other phenotypic group, 
$k\in(16,32]$.
A similar rule applies to females with $k\in(16,32]$, with the proviso that now the female picks as mate, among $N_m$ males of the same phenotypic group, the one with the largest phenotype value $k$.
The females with $k=16$ mate only with males of phenotype $k=16$.
If $Y=0.5$, the female population is not selective in mating; 
that is, panmictic mating is the behaviour of the population. 
For $Y=1.0$, the female population has a completely assortative mating behaviour.

\section{Results}

We present now simulation results for which the values of the parameters related to the first pairs 
of bitstring of the individuals' genomes were chosen to be: $A=10$, $b=5$ and $M=1$. 
The specialist populations have phenotypes $k\in[0,n_1=13]$ and $k\in[n_2=19,32]$, 
Eq. (\ref{fatorall}) and each female chooses among $N_m=5$ males.
We start the simulations with all bitstrings randomly filled
with zeroes and ones. The initial populations typically consist of 60000
individuals, half males and half females. The equilibrium population sizes
depend on the carrying capacity, but are never smaller than 38000
individuals.
\subsection{Simulations for {\it A. Labiatus} species}

\begin{figure}[htbp]
\begin{center}
\includegraphics[width=5.2cm,angle=270]{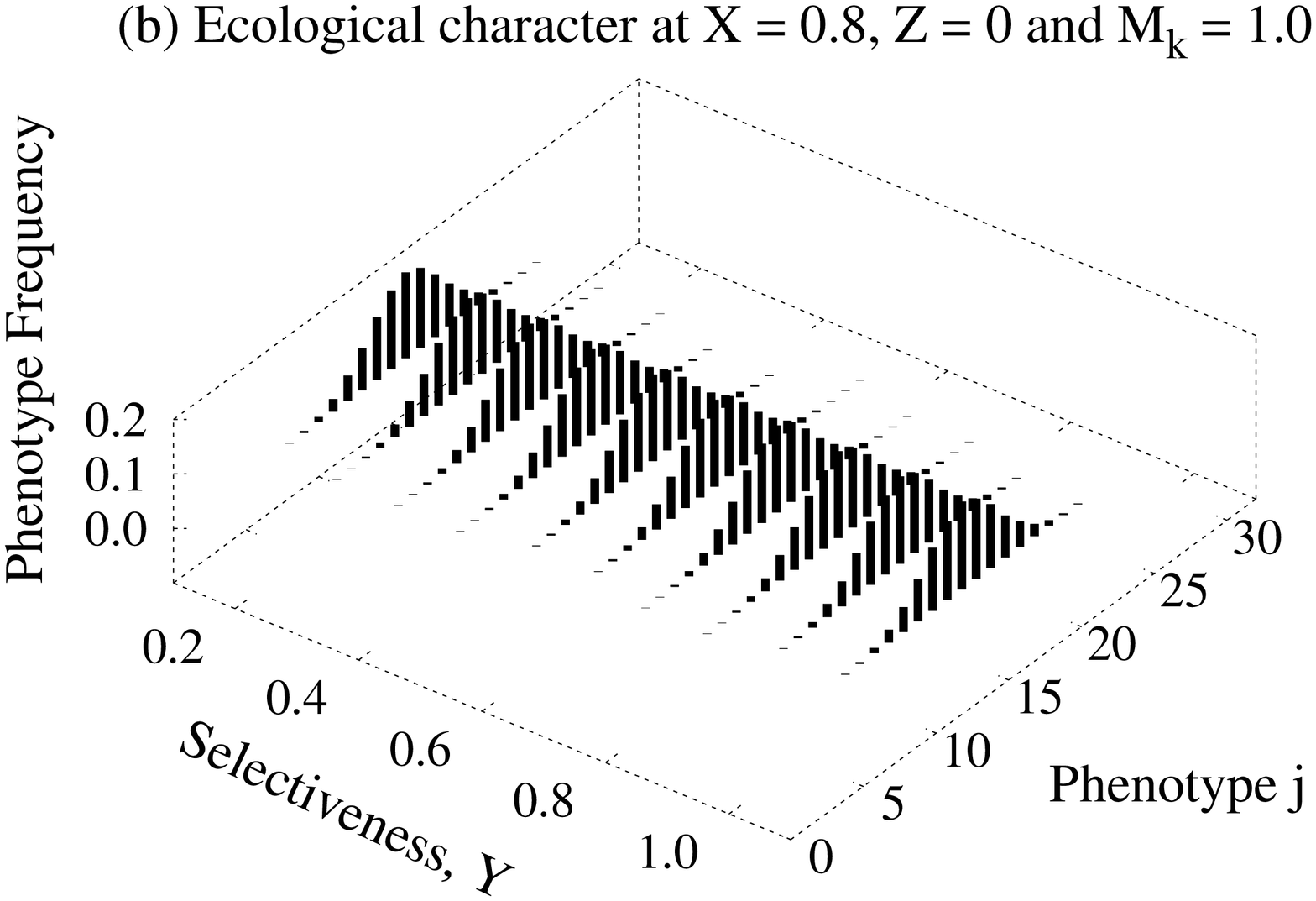}
\includegraphics[width=5.2cm,angle=270]{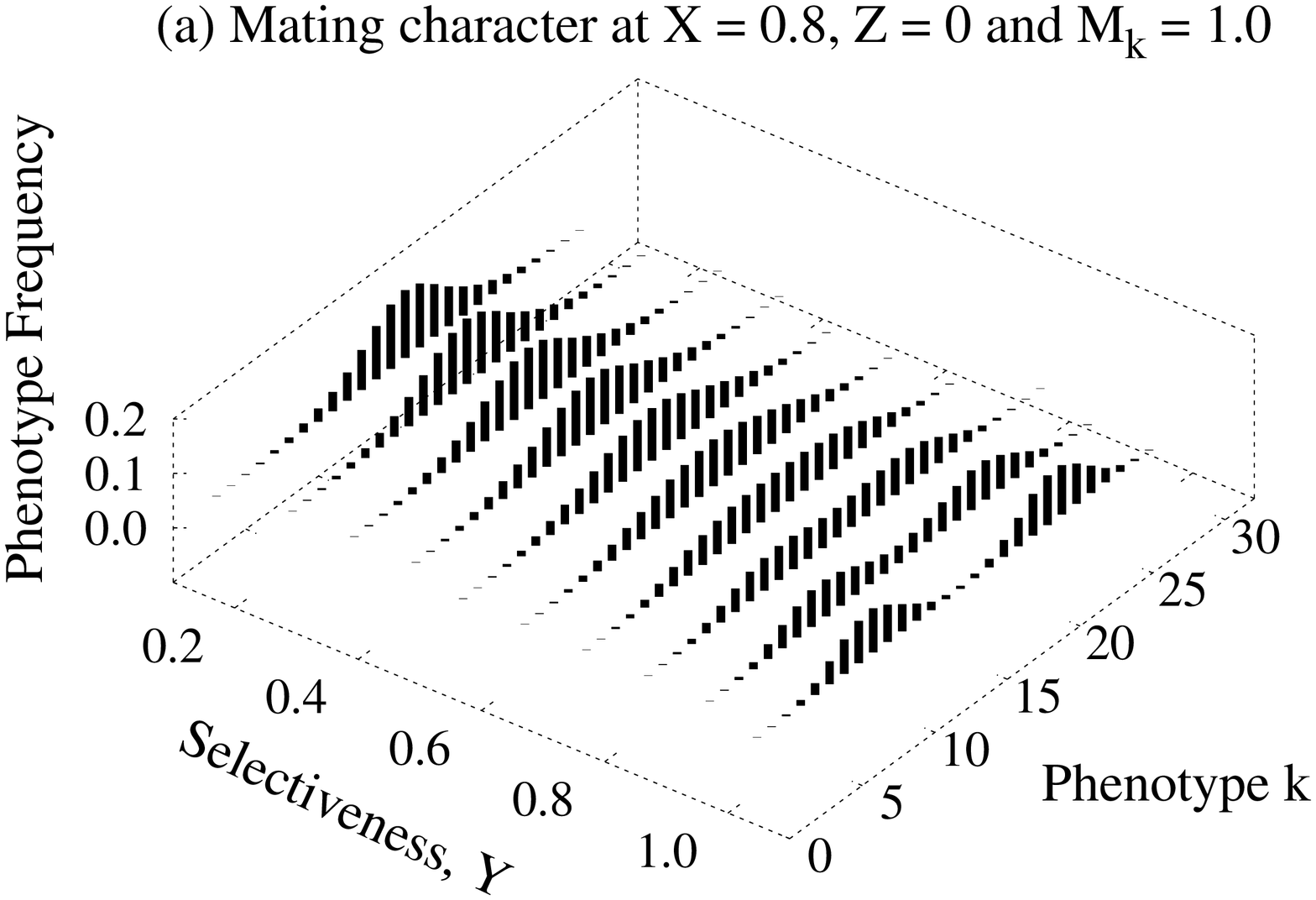}
\end{center}
\caption{\it Results for the model without disruptive selection and with 
a competition strength $X=0.8$. The phenotypes frequencies, for different 
sexual selection strengths, were measured during the last $10^6$ simulation steps.
(a) distribution of the phenotype, $j$, related to the ecological trait 
and (b) distribution of the phenotype, $k$, related to the mating trait.}
\label{fig:labiatus}
\end{figure}

We took $Z=0$ in the resource distribution, $F(j,t)$ of Eq. (\ref{eq:gauss}), 
and chose for the mutation probability per locus of the ecological and for the mating trait 
the values $M_j=1.0$ and $M_{k}=1.0$, respectively.
For all values of the sexual selection strength, $Y$, the phenotype frequency of the ecological trait 
is a stationary gaussian distribution, Fig. \ref{fig:labiatus} (a).
The phenotype frequency of the mating trait is a bimodal distribution 
for $Y>0.7$ and a gaussian distribution, with a mean value $k=16$, 
for the other $Y$ values, Fig. \ref{fig:labiatus} (b).

\begin{figure}[htbp]
\begin{center}
\includegraphics[width=5.2cm,angle=270]{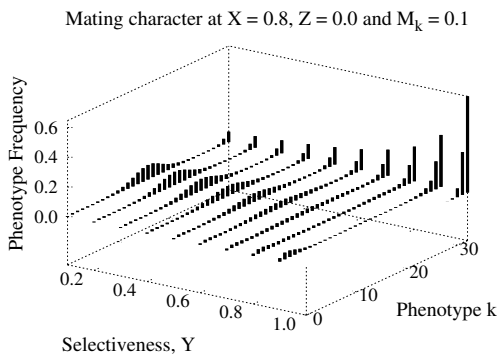}
\end{center}
\caption{\it The phenotypes frequencies of the mating trait for different 
sexual selection strength, $Y$, without disruptive natural selection and 
at a small mutation probability of the mating trait, $M_k$.}
\label{fig:labiatusM2}
\end{figure}

All these results, as well as the next one shown, are valid for all values of the 
competition strength smaller than one, $X<1$.  
If we change $N_m=5$ to smaller values, the phenotype frequency related to 
the mating trait is no longer a stationary distribution. 
If the mutation probability per locus of the ecological trait, $M_j$, is 
smaller than $1.0$, for example $M_j=0.1$, the distributions in 
Fig. \ref{fig:labiatus} (a) and (b) do not change. 
When the mutation probability of the mating trait, $M_k$, changes from $1.0$
to $0.1$, for example, the phenotype frequency of the ecological trait, 
Fig. \ref{fig:labiatus} (a), does not change either.
However, for $M_{k}=0.1$ and $Y>0.5$ the phenotype frequency of the 
mating trait, Fig. \ref{fig:labiatus} (b), changes to a unimodal distribution  
peaked at $k=0$ or $k=32$, with a $0.5$ probability for each.
For example, the phenotype frequency of the mating trait, for $Y>0.5$, is 
a distribution with a peak at $k=32$, while for other values of $Y$ and 
still $M_{k}=0.1$ the distribution is bimodal peaking at $k=16$ and 
$k=32$, Fig. \ref{fig:labiatusM2}. 
That means, if the mutation probability of the mating trait, $M_{k}$, 
has small values, the population suffers a genetic drift
which becomes faster as the value of $Y$ becomes larger, meaning that the 
behaviour of the female population is predominantly one of assortative mating. 
In other words, for strong assortative mating the disruption of the mating trait is not 
favoured for small values of the mutation probability of this trait when disruptive 
natural selection, caused by resources distributions, is not present in the population.

In what follows, the character determining sexual selection in the population 
is the colour of the individuals and the character that determines natural 
selection is the individuals' jaws morphology, which is the case in the Midas
cichlid species complex.
In common with {\it A. Labiatus} species, Fig.\ref{fig:conjuntos}, it is possible to find polychromatism 
and monomorphism in the ecological character for all values of the 
asymmetrical competition strength between the intermediate and specialist phenotypes smaller than one, 
$X<1$ in Eq. (\ref{intermediate}) and (\ref{extreme}), provided the following conditions are met: 
(i) no disruptive natural selection $Z=0$, (ii) a sexual selection strength $Y>0.7$ in the population,
and (iii) a large value for the mutation probability of the mating character $M_k=1$, 
Fig. \ref{fig:labiatus} (b).

\subsection{Simulations for {\it A. Zaliosus} species}

The characteristics of the {\it A. Zaliosus} species is to have a unimodal 
distribution for the mating trait and a bimodal distribution for the ecological one.
We have already seen, in the previous section, that for $Z=0$ there is no 
splitting of the ecological trait, unless if $Y=1.0$ and $X=1.0$. 
On the other side, for $Y=1.0$ and $X=1.0$ the mating trait is also splitted,
which is not the case of {\it A. Zaliosus} species.
So in order to simulate the {\it A. Zaliosus} process of speciation, we will 
take $Z>0$ and will first study each trait separately. 

\begin{figure}[htbp]
\begin{center}
\includegraphics[width=5.2cm,angle=270]{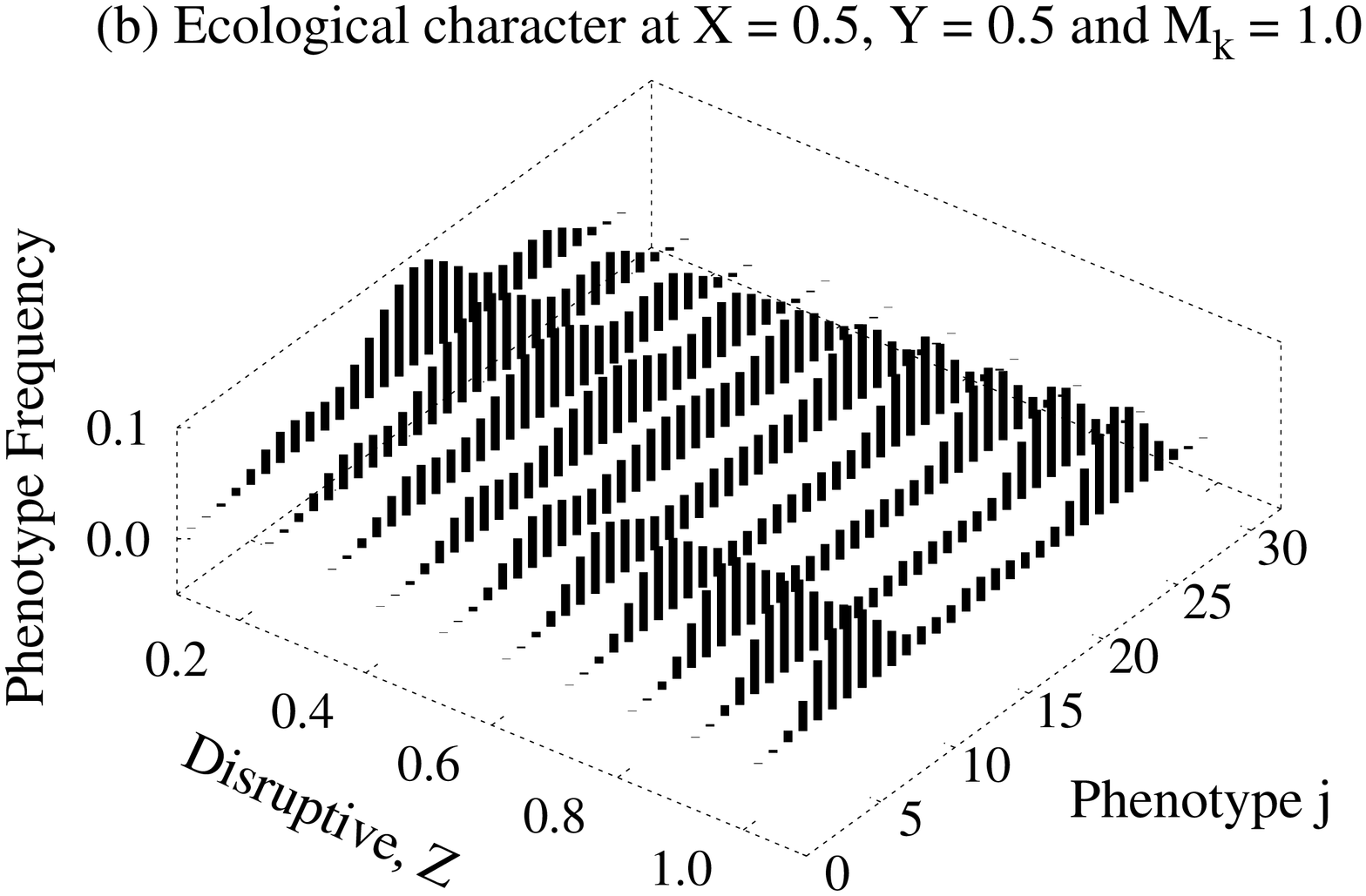}
\includegraphics[width=5.2cm,angle=270]{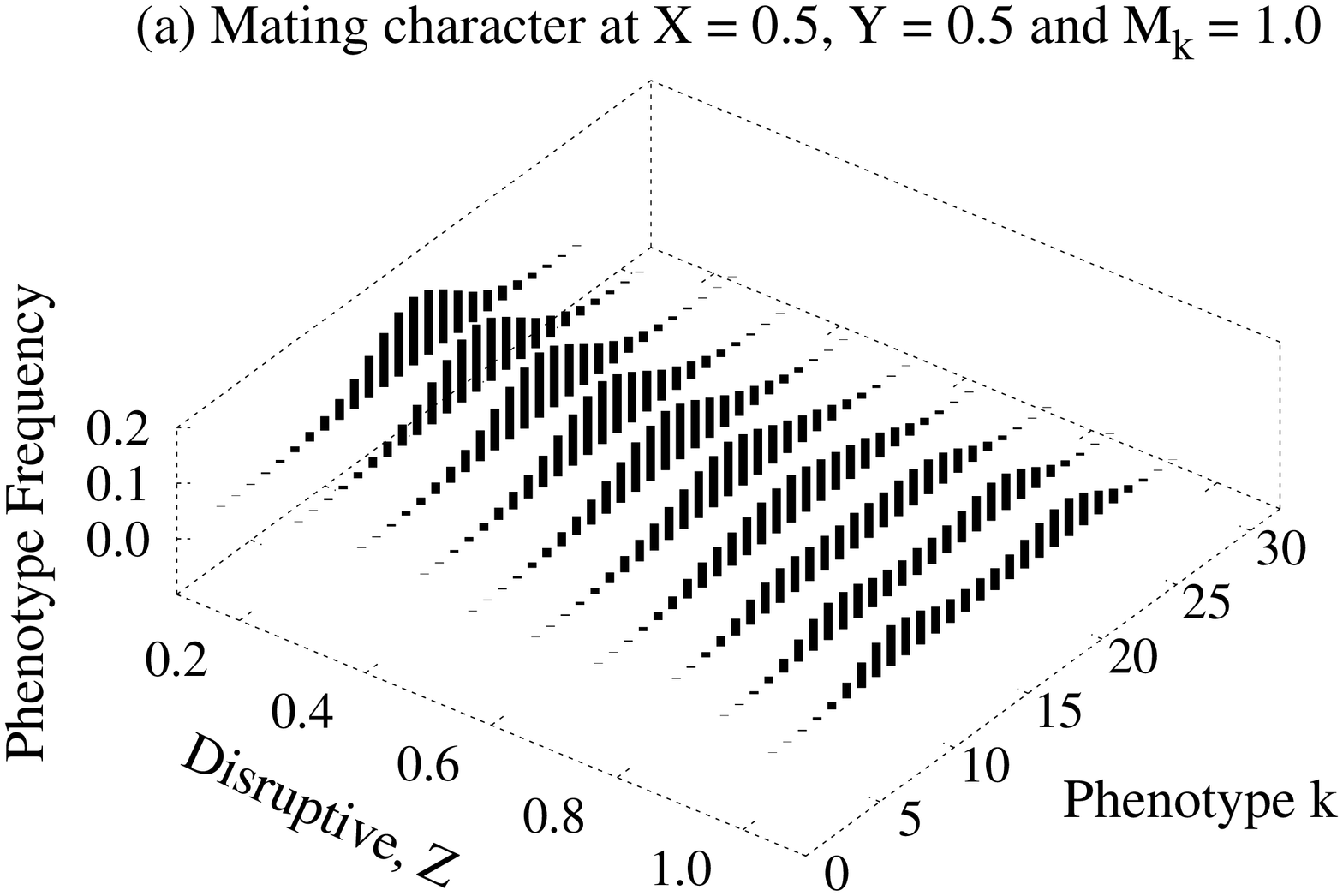}
\end{center}
\caption{\it The phenotypes frequencies for different resources 
distributions, $Z$, for $Y=0.5$ and $X=0.5$, (a) of the ecological 
trait and (b) of the mating trait.}
\label{fig:zaliosus}
\end{figure} 

Fig. \ref{fig:zaliosus} (a) shows the distribution of the ecological trait 
for $Y=0.5$ and $X=0.5$. From this figure we see that this distribution changes
from a unimodal one  to a bimodal distribution depending on the value of the 
disruptive natural selection strength, $Z$.
For $Z\approx1$, the existing intermediate phenotypes belong to individuals that 
die before reaching the minimum reproductive age.
Fig. \ref{fig:zaliosus} (b) shows the distribution of the mating trait, also
for $Y=0.5$ and $X=0.5$. It can be seen that the distribution is unimodal only 
for $Z<1$; for $Z\approx1$ it is almost bimodal but not completely, since the 
intermediate population, $k=16$, is appreciable. 
The existence of the intermediate phenotypes is due to the panmictic behaviour 
of the population, $Y=0.5$

\begin{figure}[htbp]
\begin{center}
\includegraphics[width=5.2cm,angle=270]{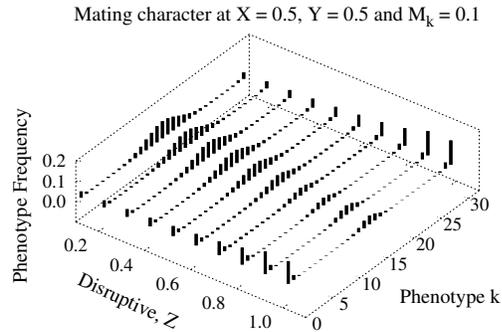}
\end{center}
\caption{\it The phenotype frequencies of the mating trait for different 
resources distributions, $Z$, and at a small mutation probability 
of the mating trait, $M_k$.}
\label{fig:zaliosusM2}
\end{figure}

The phenotype frequency of the ecological trait, Fig. \ref{fig:zaliosus} (a),
does not change if we vary the mutation probabilities $0.1\leq M_j\leq1.0$ and 
$0.1\leq M_k\leq1.0$. The phenotype frequency of the mating trait also does not
change for $0.1\leq M_j\leq1.0$. The same is not true when $M_k<1.0$, since then 
the distribution is trimodal, Fig. \ref{fig:zaliosusM2}, 
with peaks at $k=0$, $k=16$ and $k=32$ when $M_k=0.1$.

From these results we may conclude that when $Z\neq0$, the splitting of the mating
trait is favoured by small mutation probabilities, $M_k$, of this trait, which is 
not what we want for the {\it A. Zaliosus} species. 
From Figs. \ref{fig:zaliosus} (a) and (b) we can see that the proper regions of parameters to simulate the speciation process of this species are $Z>0.4$, bimodal distribution for the ecological trait, and $Z<0.8$, unimodal one for the mating trait.

\subsection{Simulations for {\it A. Citrinellus} species}

\begin{figure}[htbp]
\begin{center}
\includegraphics[width=5.2cm,angle=270]{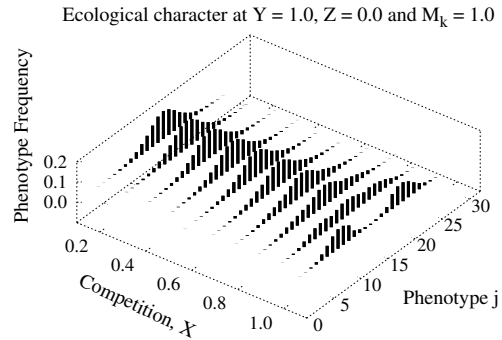}
\end{center}
\caption{\it The phenotypes frequencies of the ecological trait in a case
without disruptive natural selection and for strong sexual 
selection, $Y=1.0$.}
\label{fig:citrinellus1}
\end{figure}

One way to obtain the polycromatism and polymorphism characteristic of the
{\it A. Citrinellus} species, bimodal distributions of the both traits, is to
consider uniform distribution of resources without disruptive natural selection.
We have already shown that simulations with $Z=0.0$ and $Y=1.0$, only assortative
mating, give a bimodal distribution for the mating trait, Fig. \ref{fig:labiatus} 
(b), independently of the competition strength, $X$, provided the mutation probability is $M_k=1.0$.   
However, to obtain also bimodal distribution of the ecological trait, without 
disruptive natural selection, it is necessary to have $X=1.0$, that is, a symmetric 
competition between specialist and intermediate phenotypes, as shown in 
Fig. \ref{fig:citrinellus1}.

For $Z>0$, it is also possible to split both phenotype distributions, but then 
it is necessary to consider small mutation probabilities of the mating trait, $M_k$,
a weak assortative mating, $Y\gtrsim0.5$, and proper values of the resource distribution, $Z$, depending on the competition strength, $X$. For instance, $Z>0.7$ for $X=0.5$, Fig \ref{fig:zaliosus} (a).

\section{Discussion}

Although competition for resources and natural disruptive selection appear together in
Eq. (\ref{extreme}) and (\ref{intermediate}), they lead to rather different situations. 
While competition depends on the population sizes and affects equally individuals of the
same phenotypic group, disruptive natural selection caused by resources distributions 
acts on each particular individual, according to its phenotype. As a result, disruptive
selection becomes more effective than competition. For instance, even for a panmictic 
behaviour, disruptive natural selection may lead to a splitting of both traits, 
depending only on the values of $X$ and $Z$, as already pointed in \cite{dd}. 
A small mutation probability of the mating trait also favours this double splitting, 
as obtained in \cite{kk}.

On the other hand, in order to split both phenotypic distributions without disruptive
natural selection, $Z=0$, it is imperative to have sexual selection, and now the
splitting process depends on the values of $Y$ and $X$, and in this case the 
mutation of the mating trait must be large in order to prevent genetic 
drift effects. 

Anyway, the {\it A. Citrinellus} case is the evolution of the splitting in both 
traits distributions must be driven by disruptive natural selection or by sexual selection,
and the only statistical difference we found between these two scenarios is that the
correlation between the traits is smaller when the splitting is driven by
sexual selection. It so happens because the large mutation rate of the mating
trait, mentioned above, introduces large fluctuations in this correlation.

\begin{acknowledgments} 
We thank the agencies CNPq, FAPERJ (E-26/170.699/2004), and CAPES for
financial support and J. Nogales for fruitful discussions. 
\end{acknowledgments}

\end{document}